\documentclass[prl,aps,twocolumn,floatfix,superscriptaddress,preprintnumbers]{revtex4-1}
\usepackage{lmodern}
\usepackage{hyperref}
\hypersetup{colorlinks=true,linkcolor=purple,anchorcolor=blue,citecolor=blue, filecolor=blue,urlcolor=blue,bookmarksnumbered=true,
pdfview=FitB
}
\usepackage{mathrsfs}
\usepackage{rsfso}
\usepackage{color}
\usepackage{xcolor}
\usepackage{ulem}
\usepackage{csquotes}
\colorlet{purple1}{blue!70!red}
\colorlet{darkred}{red!50!black}

\usepackage{graphicx}
\usepackage[bf,SL,BF]{subfigure}
\usepackage{psfrag}
\usepackage{color}
\usepackage{amssymb}
\usepackage{amsmath}
\usepackage{epstopdf}
\usepackage{natbib}
\newcommand{\be}{\begin{eqnarray}}
\newcommand{\ee}{\end{eqnarray}}

\begin{document}

\title{Polarized gluon distribution in the proton from
holographic light-front QCD}

\author{Bheemsehan~Gurjar}
\email{gbheem@iitk.ac.in} 
\affiliation{Indian Institute of Technology Kanpur, Kanpur-208016, India}

\author{Chandan~Mondal}
\email{mondal@impcas.ac.cn} 
\affiliation{Institute of Modern Physics, Chinese Academy of Sciences, Lanzhou 730000, China}
\affiliation{School of Nuclear Science and Technology, University of Chinese Academy of Sciences, Beijing 100049, China}

\author{Dipankar~Chakrabarti}
\email{dipankar@iitk.ac.in} 
\affiliation{Indian Institute of Technology Kanpur, Kanpur-208016, India}

\date{\today}

\begin{abstract}
We obtain the gluon parton distribution functions (PDFs) in the proton within the extended light-front holographic QCD framework, where the proton couples with the spin-two Pomeron in Anti-de Sitter space, together with constraints imposed by the Veneziano model. The gluon helicity asymmetry, after satisfying the perturbative QCD constraints at small and large longitudinal momentum regions, agrees with existing experimental measurements. The polarized gluon distribution is consistent with global analyses. We predict the gluon helicity contribution to the proton spin, $\Delta G=0.221^{+0.056}_{-0.044}$, close to the recent analysis with updated data sets and PHENIX
measurement and the lattice QCD simulations. We subsequently present the unpolarized and polarized gluon generalized parton distributions in the proton.

\end{abstract}

\maketitle

\section{Introduction}
How the proton spin emerges from its constituents, quarks and gluons, is one of the key puzzles in modern particle and nuclear physics. In this context, the proton spin decomposition into separate quark and gluon contributions is not unique and intrinsically debatable due to quark-gluon couplings~\cite{Leader:2021pqf,Wakamatsu:2014zza,Liu:2015xha}.
The well-known proton spin decomposition proposed by Jaffe and Manohar reads~\cite{Jaffe:1989jz},
\begin{equation}\label{spin_sum}
\frac{1}{2}=\frac{1}{2}\Delta \Sigma +L_{q}+L_{g}+\Delta G\,,
\end{equation}
with quark helicity $\frac{1}{2}\Delta\Sigma$, quark orbital angular momentum (OAM) $L_{q}$, gluon helicity $\Delta G$, and gluon OAM $L_{g}$. The quark and gluon helicity components are related to their polarized parton distributions functions (PDFs), while their OAM contributions are linked to the generalized parton distributions (GPDs)~\cite{Ji:1996nm,Diehl:2003ny,Belitsky:2005qn,Goeke:2001tz}. The Jaffe-Manohar decomposition is not the unique way to decompose the proton spin. Ji proposed a frame-independent and gauge-invariant approach for dividing the proton spin into quark helicity, quark OAM, and gluon total angular momentum contributions~\cite{Ji:1996ek}. On the basis of naive understanding from the quark model, one would expect that the quark spin component contributes the majority of the spin sum rules. However, the famous European Muon
Collaboration (EMC) experiment~\cite{EuropeanMuon:1987isl} demonstrated that only a tiny portion of the proton spin, $\Delta\Sigma=0.060(47)(69)$ at $Q^{2}=10$ GeV$^{2}$~\cite{EuropeanMuon:1987isl,EuropeanMuon:1989yki}, is contributed by the quark spin, that triggered the problem of so-called `proton spin puzzle'. After a substantial amount of research over the last several decades, it has now been determined that the quark helicity component contributes just around 30$\%$ to the proton spin~\cite{deFlorian:2009vb,Nocera:2014gqa,Ethier:2017zbq}. 

The gluon distributions are extracted less precisely than the quark distributions. However, the accuracy of extracted unpolarized gluon distribution $g(x)$ has been greatly enhanced over the last decade and there are still improvements to be made, specifically at small-$x$ region. 
In contrast to the unpolarized gluon PDF, the polarized gluon PDF $\Delta g(x)$ is poorly known. It has been shown in Ref.~\cite{deFlorian:2014yva} that $\Delta g(x)$ is positive and nonzero in the momentum fraction range: $0.05< x<0.2$. However, the distribution is quite ambiguous, particularly in the small $x$-region. For a recent review, see Ref.~\cite{Ji:2020ena}. Fortunately, the upcoming Electron-Ion-Collider (EIC)~\cite{Accardi:2012qut} aims to accurately determine $\Delta g(x)$ at low-$x$ and provides rigorous limits on the polarized gluon distribution.

We compute the polarized gluon distribution within the framework based on holographic light-front QCD (HLFQCD)~\cite{Brodsky:2014yha} and the generalized Veneziano model~\cite{Veneziano:1968yb,Ademollo:1969wd,Landshoff:1970ce}. HLFQCD is a nonperturbative approach based on the gauge-gravity correspondence~\cite{Maldacena:1997re} and its holographic mapping on light-front QCD~\cite{Brodsky:2006uqa,deTeramond:2008ht}. A remarkable feature of HLFQCD is that it reproduces the hadronic spectra with least number of parameters, the confining strength and the effective quark masses. The effective confining potential for the QCD bound-states is uniquely determined by an underlying superconformal algebra~\cite{Fubini:1984hf,deTeramond:2014asa,Dosch:2015nwa}. HLFQCD generates the structure of hadronic spectra as anticipated by dual models, most notably the Veneziano model~\cite{Veneziano:1968yb,Ademollo:1969wd,Landshoff:1970ce} with its defining characteristics, linear Regge trajectories with a universal slope. 
 This novel approach has been successfully employed to simultaneously derive the quark distributions in the nucleon and the pion~\cite{deTeramond:2018ecg,Liu:2019vsn} as well as the
strange-antistrange and the charm-anticharm asymmetries in the nucleon~\cite{Sufian:2018cpj,Sufian:2020coz}. Recently, the unpolarized gluon distributions in the nucleon and the pion have also been successfully determined using the universality properties of parton distributions in LFHQCD~\cite{deTeramond:2021lxc}.

We determine the polarized gluon distribution $\Delta g(x)$ and the gluon helicity asymmetry $\Delta g(x)/g(x)$ as well as the gluon GPDs in the proton. 
One salient issue can be addressed with our study, which  concerns the description of the experimental data on the gluon helicity contribution $\Delta G$ to the proton spin sum rule, Eq.~\eqref{spin_sum}.
The RHIC spin program at BNL~\cite{STAR:2014wox,deFlorian:2014yva,Nocera:2014gqa,Ethier:2017zbq,STAR:2021mqa} and the recent lattice QCD simulations~\cite{Yang:2016plb} have revealed that $\Delta G=\int_{0}^{1}{\rm d}x\,\Delta g(x)$ is nonvanishing and likely sizable. Several global analyses have been performed to establish limitations on $\Delta G$ using various experimental data sets and parametrizations~\cite{Gehrmann:1995ag,Gluck:2000dy,Blumlein:2002qeu,Leader:2005ci}. Using updated data sets and PHENIX measurement~\cite{PHENIX:2008swq}, a recent extraction yielded
$\Delta G= 0.2$ with a restriction of $-0.7<\Delta G<0.5$ for the gluon momentum fraction $0.02\leq x \leq 0.3$. It was reported in Ref.~\cite{Nocera:2014gqa} that $\Delta G=\int_{0.05}^{0.2}{\rm d}x\,\Delta g(x)=0.23(6)$ and in Ref.~\cite{deFlorian:2014yva}, $\Delta G=\int_{0.05}^{1}{\rm d}x\,\Delta g(x)=0.19(6)$. Meanwhile, the large-momentum effective theory~\cite{Ji:2013dva,Ji:2014gla} provides $\Delta G(\mu^{2}=10$ GeV$^2)= 0.251(47)(16)$, which is almost half of the proton. In order to confine $\Delta g(x)$ at low-$x$,  some theoretical constraints have been discussed in Ref.~\cite{Kovchegov:2017lsr}. Several experiments are now being conducted at the RHIC~\cite{PHENIX:2014gbf,Bunce:2000uv}, HERMES~\cite{HERMES:2008abz}, JLab~\cite{Dudek:2012vr}, COMPASS~\cite{COMPASS:2018pup} to obtain high-precision measurements of the gluon helicity $\Delta G$. Addressing this fundamental issue demands a unified framework, such as we demonstrate here, that adequately provides a prediction of the expected data for the gluon helicity from the future experiments.

\section{Gluon distribution functions} \label{GFFsproton}
\subsection{Unpolarized PDF}
The unpolarized gluon distribution function can be derived from the HLFQCD expression of its gravitational form factor (GFF)~\cite{deTeramond:2021lxc}.
To compute the gluon GFF for arbitrary twist-$\tau$ Fock state in the light-front Fock expansion of the
proton state, $A^g_\tau(t)$, the Pomeron is considered to couple mainly to the constituent gluon~\cite{STAR:2012fiw,Ewerz:2013kda,Ewerz:2016onn,Lebiedowicz:2016zka,Britzger:2019lvc}.  The lowest twist is the $\tau = 4$ Fock state $|uudg\rangle$ in the proton containing a dynamical gluon. The Pomeron couples to the dynamical gluon over a distance $\sim 1 / \sqrt{\alpha^\prime_P}$, where $\alpha^\prime_P$ defines the slope of the effective Regge trajectory of the Pomeron. In
LFHQCD, the GFF $A^g_\tau(t)$ can be expressed in terms of the Euler Beta function $B(u\,,v)$ as~\cite{deTeramond:2021lxc}
\be \label{GFFA}
A^{g}_{\tau}(t)=\frac{1}{N_{\tau}}B( \tau -1,2-\alpha_{P}(t)),
\ee
where $N_\tau =  B \left(\tau - 1, 2 - \alpha_P(0)\right)$ and 
\be\label{Regge_pomeron}
\alpha_{P}(t) = \alpha_{ P}(0) + \alpha'_{P} t\,,
\ee
is the the effective Regge trajectory of the Donnachie and Landshoff's soft pomeron~\cite{Donnachie:1992ny} with  intercept $\alpha_P(0) \simeq 1.08$,  slope $\alpha^\prime_P \simeq  0.25 \, {\rm GeV}^{-2}$~\cite{ParticleDataGroup:2020ssz}, and $t=-Q^2$ is the
square of transferred momentum. Eq.~\eqref{GFFA} has the same structure as a generalization of the Veneziano amplitude~\cite{Veneziano:1968yb, Ademollo:1969wd, Landshoff:1970ce} for a spin-two current. Note that while writing Eq.~\eqref{GFFA}, only the dilaton profile~\cite{Karch:2006pv}: $e^{\varphi_g (z)} = e^{- \lambda_g z^2}$ with $\lambda_g = 1/4  \alpha_P^\prime \simeq 1 \, {\rm GeV}^2$ describing Pomeron exchange has been considered. This sets the scale when computing the gluon GFFs and GPDs. Pomeron exchange is recognized as
the graviton of the dual AdS theory~\cite{Brower:2006ea, Cornalba:2008sp, Domokos:2009hm, Brower:2010wf, Costa:2012fw, Costa:2013uia, Amorim:2021gat}. Meanwhile, only the dilaton corresponding to Reggeon exchange: $e^{\varphi_q (z)}= e^{\lambda_q z^2}$ with $\lambda_q = 1/4 \alpha_\rho^\prime  \simeq (0.5 \, {\rm GeV})^2$, needs to be assumed when deriving the electromagnetic form factors and quark GPDs~\cite{deTeramond:2018ecg}.

Using the integral representation of the Euler Beta function,
\be\label{betafun}
B(u,v)=B(v,u)
=\int_{0}^{1}{\rm d}y\,y^{u-1}(1-y)^{v-1},
\ee
where $ \Re(u)>0$ and $\Re (v)>0$, in Eq.~\eqref{GFFA}, the gluon GFF $A_\tau(t)$ can be recast in the reparametrization invariant form as
\begin{align}  \label{GFF1}
A_{\tau}^{g}(t)=\frac{1}{N_{\tau}}\int_{0}^{1} {\rm d}x\, w^{\prime}(x)w(x)^{1-\alpha_{P}(t)}[1-w(x)]^{\tau-2},
\end{align}
provided that $w(x)$ is a monotonically increasing function and satisfies the constraints $w(0) = 0$,  $w(1) = 1$  and $w'(x) \ge 0$ with $x\in [0,\,1]$. The reparametrization function  $w(x)$ is introduced in Ref.~\cite{deTeramond:2018ecg, Liu:2019vsn,deTeramond:2021lxc} is given by
\be \label{UniversalFun}
w(x)=x^{1-x}e^{-a(1-x)^{2}}\,,\label{rep_fn}
\ee
with the parameter $a=0.48$ $\pm$ 0.04. 
The gluon GFF $A_{\tau}^{g}(t)$ can also be written as the first moment of the gluon GPD at zero skewness, $H^{g}_{\tau}(x,t)$,
\be \label{GPDH}
 A^{g}_{\tau}(t)=\int_{0}^{1}{\rm d}x\,xH_{\tau}^{g}(x,t)
 =\int_{0}^{1}{\rm d}x\,xg_{\tau}(x)\,e^{tf(x)}\,,
\ee
where $g_{\tau}(x)$ is the collinear unpolarized gluon PDF of twist-$\tau$ and $f(x)$ is the profile function. Comparing Eq.~\eqref{GPDH} with Eq.~\eqref{GFF1}, one can extract both functions, $g_{\tau}(x)$ and $f(x)$, in terms of the universal reparametrization function $w(x)$, 
\begin{align} \label{gtaux}
g_{\tau}(x)&=\frac{1}{N_{\tau}}\frac{w^{\prime}(x)}{x}[1-w(x)]^{\tau-2}w(x)^{1-\alpha_{P}(0)}\,,\\
\label{profilefun}
f(x)&=\alpha^{\prime}_{P}\log \bigg( \frac{1}{w(x)} \bigg)\,, 
\end{align}
with the normalization condition $\int_{0}^{1}dx x g_{\tau}(x)=1$. The PDF for the gluon in the proton is written as the sum of contributions from all Fock states i.e., $g(x)=\sum_{\tau} c_{\tau}g_{\tau}(x)$. Note that we only consider the leading term, $\tau=4$,  and the coefficient $c_{\tau=4}=0.225 \pm 0.014$~\cite{deTeramond:2021lxc}  has been determined by using the momentum sum rule
\begin{align}
    \int_0^1 {\rm d}x\, x \big[g(x) + \sum_{q} q(x) \big] = 1,
\end{align} 
with the help of quark distributions $q(x)$ at the hadronic scale $\mu^2 \sim 1$  GeV obtained previously within the HLFQCD framework~\cite{deTeramond:2018ecg}.

\begin{figure}
\begin{center}
\includegraphics[scale=0.55]{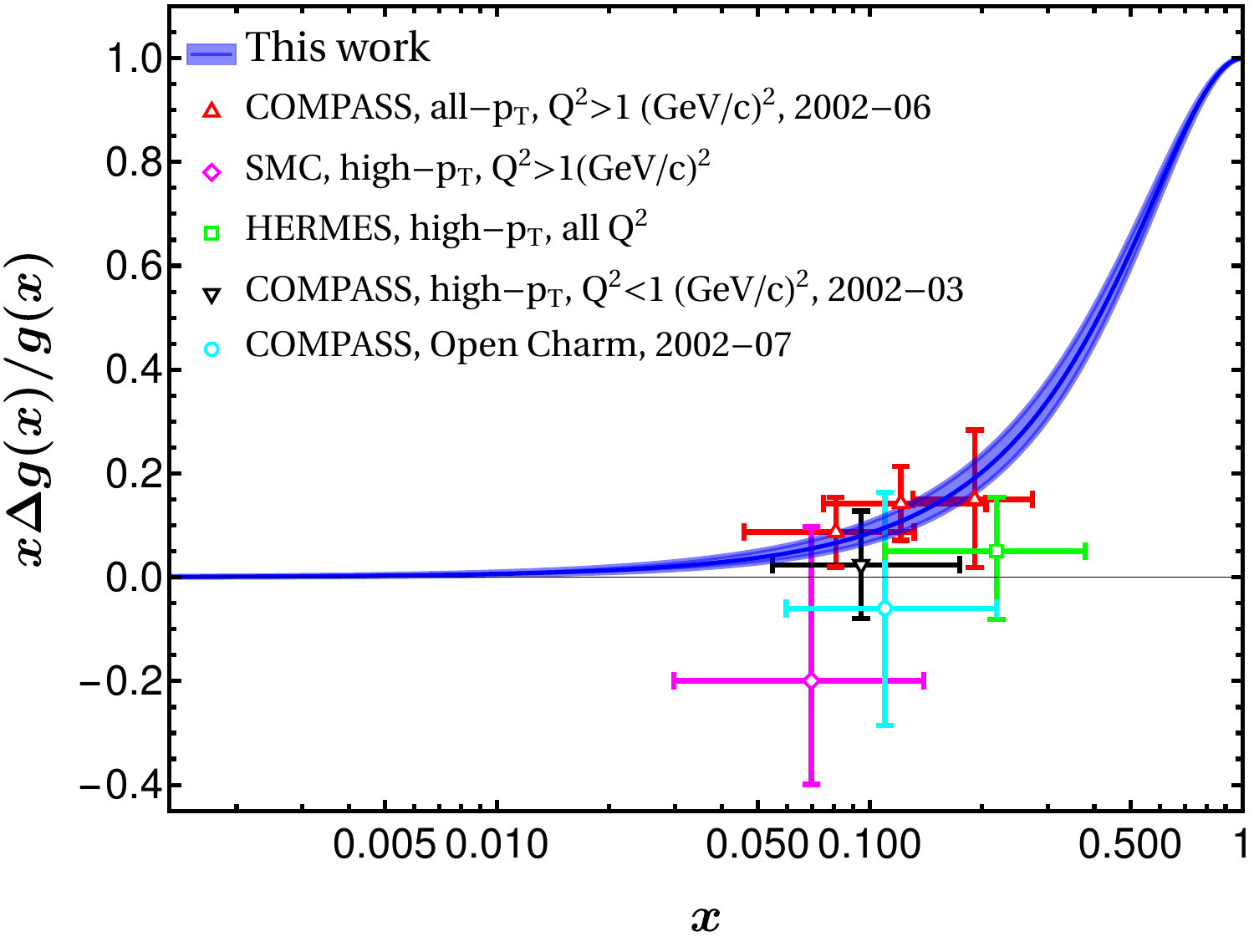}
\end{center}
\caption{The gluon helicity asymmetry, $\Delta g(x)/g(x)$, in the proton (blue band) is compared with the available experimental data~\cite{COMPASS:2005qpp,COMPASS:2015pim,HERMES:2010nas,SpinMuonSMC:2004jrx,COMPASS:2012mpe}. The direct measurements of COMPASS~\cite{COMPASS:2005qpp,COMPASS:2015pim}, HERMES~\cite{HERMES:2010nas} and SMC~\cite{SpinMuonSMC:2004jrx} are obtained in the leading order from high $p_{T}$ hadrons while open charm muon production at COMPASS~\cite{COMPASS:2012mpe} are taken from next-to-leading order at different values of $x$. The error band in our result is due to the spread in the Regge intercept $\widetilde{\alpha}_{P}(0)\equiv 0-0.16$ and  the uncertainties in the parameter $a=0.48$ $\pm$ 0.04 appearing in the reparametrization function $w(x)$, Eq.~\eqref{rep_fn}}
\label{asymmetry}
\end{figure}
 
\subsection{Helicity PDF}\label{helicitydistribution}
Polarized gluon distributions can be evaluated by using Eq.~\eqref{gtaux} but with different Pomeron Regge trajectory,
\be \label{helicitypdf}
\Delta g_{\tau}(x)=\frac{1}{N_{\tau}}\frac{w^{\prime}(x)}{x}[1-w(x)]^{\tau-2}w(x)^{1-\alpha_{P}^{\prime}(0)},
\ee
where the Regge trajectory is given by
\be \label{Reggepolarized}
\widetilde{\alpha}_{P}(t)=\widetilde{\alpha}_{P}(0)+\alpha^{\prime}_{P}t\,.
\ee
Note that the slope of the Regge trajectories is universal, while their intercepts are different for the unpolarized and the polarized gluon distributions.
We determine the value of the intercept $\widetilde{\alpha}_{P}(0)$ by requiring the result to fit the experimental data for the gluon asymmetry ratio, $\Delta g(x)/g(x)$, together with the constraints of $\Delta g(x)/g(x)$ at $x \rightarrow 1$ and $x \rightarrow 0$~\cite{Brodsky:1989db,Brodsky:1994kg}. 
In HLFQCD, the gluon helicity asymmetry behaves as
\be 
\frac{\Delta g_{\tau}(x)}{g_{\tau}(x)} = w(x)^{\alpha_{P}(0)-\widetilde{\alpha}_{P}(0)}\,,
\ee
where the exponent, $\alpha_{P}(0)-\widetilde{\alpha}_{P}(0)$, is the difference between the intercepts of unpolarized and polarized Regge trajectries. Note that any value of $\widetilde{\alpha}_{P}(0)<\alpha_{P}(0)$ satisfies the pQCD predictions for the helicity asymmetry retention~\cite{Brodsky:1989db,Brodsky:1994kg}. We fix $\widetilde{\alpha}_{P}(0)\equiv 0-0.16$ by fitting the helicity asymmetry to the experimental data. At our center value of $\widetilde{\alpha}_{P}(0)=0.08$, the $\chi^2$ per d.o.f. for the fit is $1.5$.

Figure~\ref{asymmetry} confirms that the gluon helicity asymmetry $\Delta g(x)/g(x)$ satisfies the pQCD constraints at the end points. The helicity asymmetry decreases to zero at small-$x$  and increases to one when $x$ approaches one. The model uncertainty (blue band) includes the uncertainties in the parameter $a$ appearing in the reparametrization function $w(x)$, Eq.~\eqref{rep_fn} and the spread in the Regge intercept $\widetilde{\alpha}_{P}(0)$. We compare the ratio $\Delta g(x)/g(x)$ with the data at different gluon longitudinal momentum extracted from high $p_{T}$ hadrons in the leading-order analyses~\cite{COMPASS:2005qpp,COMPASS:2015pim} and from the open charm production in the next-to-leading order analysis~\cite{COMPASS:2012mpe} at COMPASS, from high $p_{T}$ hadrons at leading-order analyses by Spin Muon Collaboration (SMC) at CERN~\cite{SpinMuonSMC:2004jrx} and at HERMES experiment~\cite{HERMES:2010nas}. We find a good agreement between our result and the COMPASS data. Note that there still remain large uncertainties of the ratio $\Delta g(x)/g(x)$, including even the sign from different experiments.

\begin{figure}
\begin{center}
\includegraphics[scale=0.31]{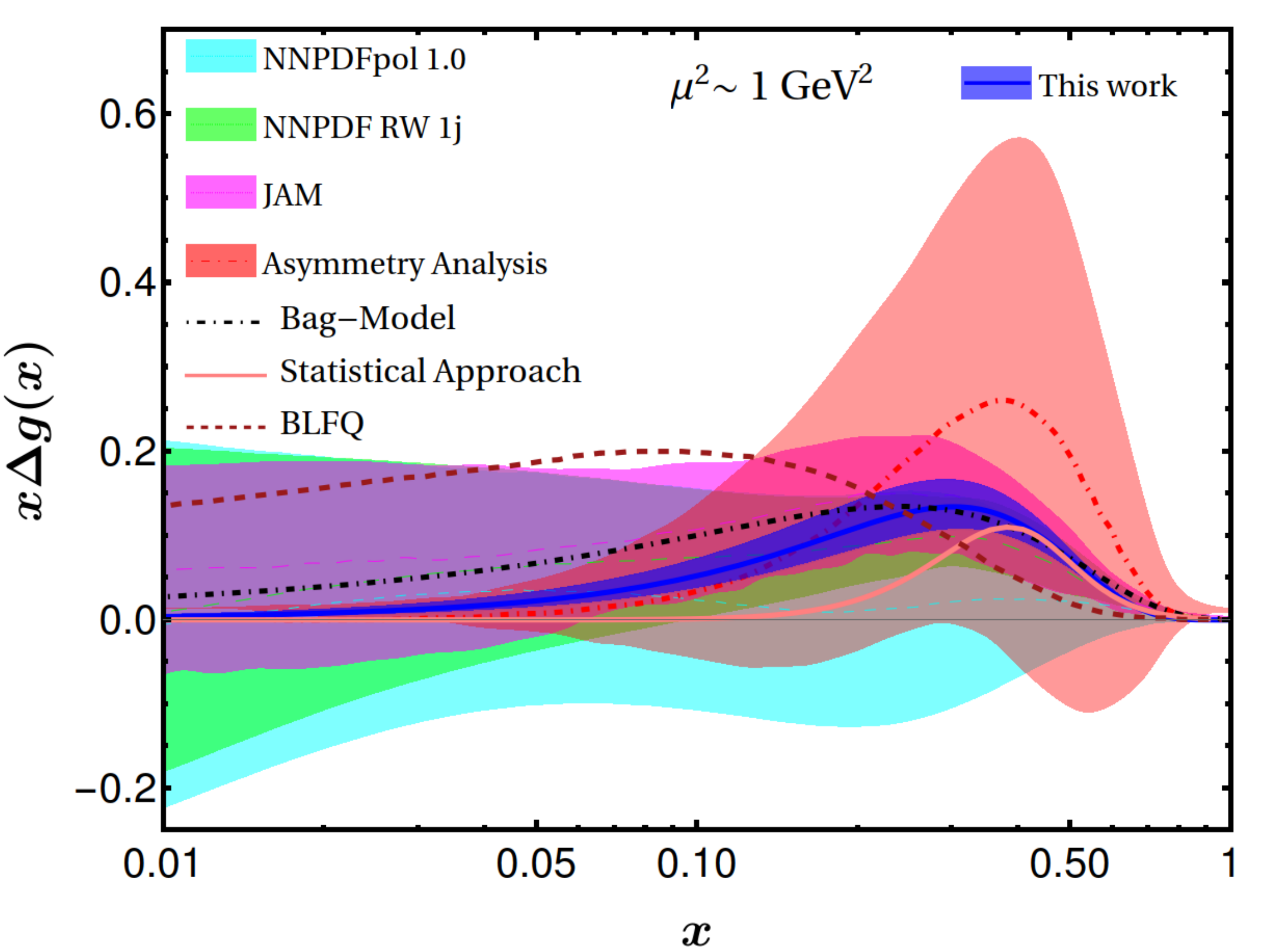}
\end{center}
\caption{The polarized gluon distribution $x\Delta g(x)$ at the scale $\mu^2\sim 1$ GeV$^2$ (blue band) is compared with the global analyses by NNPDFpol1.0~\cite{Nocera:2014gqa} (cyan band) Collaborations, the NNPDFpol1.0 reweight RHIC data~\cite{Nocera:2014gqa} (green band), and JAM~\cite{Sato:2016tuz} (magenta band) as well as with other theoretical studies: the Bag model~\cite{Chen:2006ng} (black dash-dotted line), phenomenological fit by Asymmetry Analysis Collaboration~\cite{Hirai:2003pm} (red dash-dotted line surrounded by an uncertainty band), the statistical approach~\cite{Bourrely:2014uha} (pink solid line), and the basis light-front quantization (BLFQ) approach~\cite{Xu:2022abw} (purple dashed line).}
\label{polarizedgluonPDF}
\end{figure}

Having specified the gluon helicity asymmetry ratio, we are now in a position to present explicitly the gluon helicity PDF in HLFQCD. We show the intrinsic nonperturbative gluon helicity distribution, $x\Delta g(x)$, defined at the initial scale $\mu^2 \sim 1$ in Fig.~\ref{polarizedgluonPDF}, where we compare our prediction with the global analyses by  the NNPDFpol1.0~\cite{Nocera:2014gqa} and the JAM~\cite{Sato:2016tuz} Collaborations as well as with other theoretical studies~\cite{Chen:2006ng,Hirai:2003pm,Bourrely:2014uha,Bacchetta:2020vty,Xu:2022abw}. We find a good consistency between our
prediction for the proton's gluon helicity PDF
and the global fits and the results obtained from various
theoretical approaches. The uncertainty band stems from the of model parameters, $c_{\tau=4}$, $a$, and $\widetilde{\alpha}_{P}(0)$. We notice that at the model scale, the percentage uncertainty of $x\Delta g(x)$ is larger than that of $x g(x)$.  It should be noted that there are large
uncertainties in the global analyses and thus $\Delta g(x)$ is poorly constrained, including
even the sign, especially in the small-$x$ region but also in
the large-$x$ region.

The  gluon spin contribution $\Delta G$ to the proton spin is given by the first moment of the gluon helicity PDF $\Delta g(x)$. Our current analysis predicts that the gluon spin, 
$\Delta G=0.221^{+0.056}_{-0.044}$, 
is sizeable to the proton spin and close to the recent analysis with updated data sets and PHENIX measurement~\cite{PHENIX:2008swq}, which  yielded $\Delta G=0.2$  for $x_g\in [0.02, 0.3]$. Excluding the $x_g<0.05$ region, the value of $\Delta G =0.23(6)$ for $x_g\in [0.05, 0.2]$~\cite{Nocera:2014gqa} and $\Delta G =0.19(6)$ for $x_g\in [0.05, 1]$~\cite{deFlorian:2014yva} were reported. Whereas,  the lattice QCD calculation at physical pion mass predicts $\Delta G = 0.251(47)(16)$~\cite{Yang:2016plb}. Due to the current accuracy of experimentally measured data, the phenomenological extraction of $\Delta G$ is sensitive to the parametrization form in the global analyses. One will find large uncertainties of $\Delta g(x)$ and thus, very poor
constraint on $\Delta G$, if permitting
a possible sign change of $\Delta g(x)$ at some values of $x$~\cite{Zhou:2022wzm}.  Future  measurements of $\Delta g(x)$ in the $x_g<0.02$ are necessitated to reduce the uncertainty in $\Delta G$.
 Resolving this issue is one of the major goals of the future EICs~\cite{Accardi:2012qut,AbdulKhalek:2021gbh}.

\begin{figure}
\begin{center}
\includegraphics[scale=0.47]{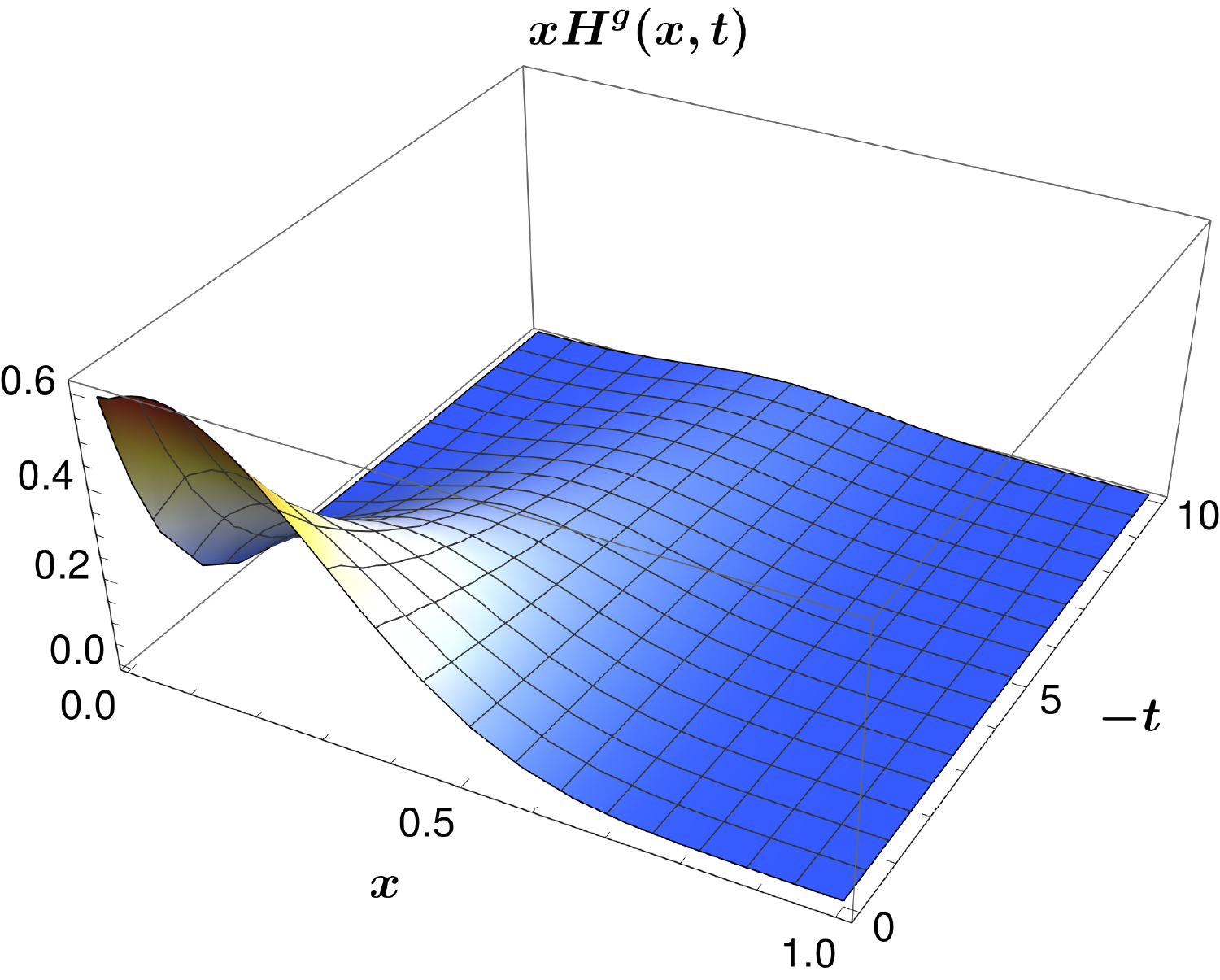}
\includegraphics[scale=0.47]{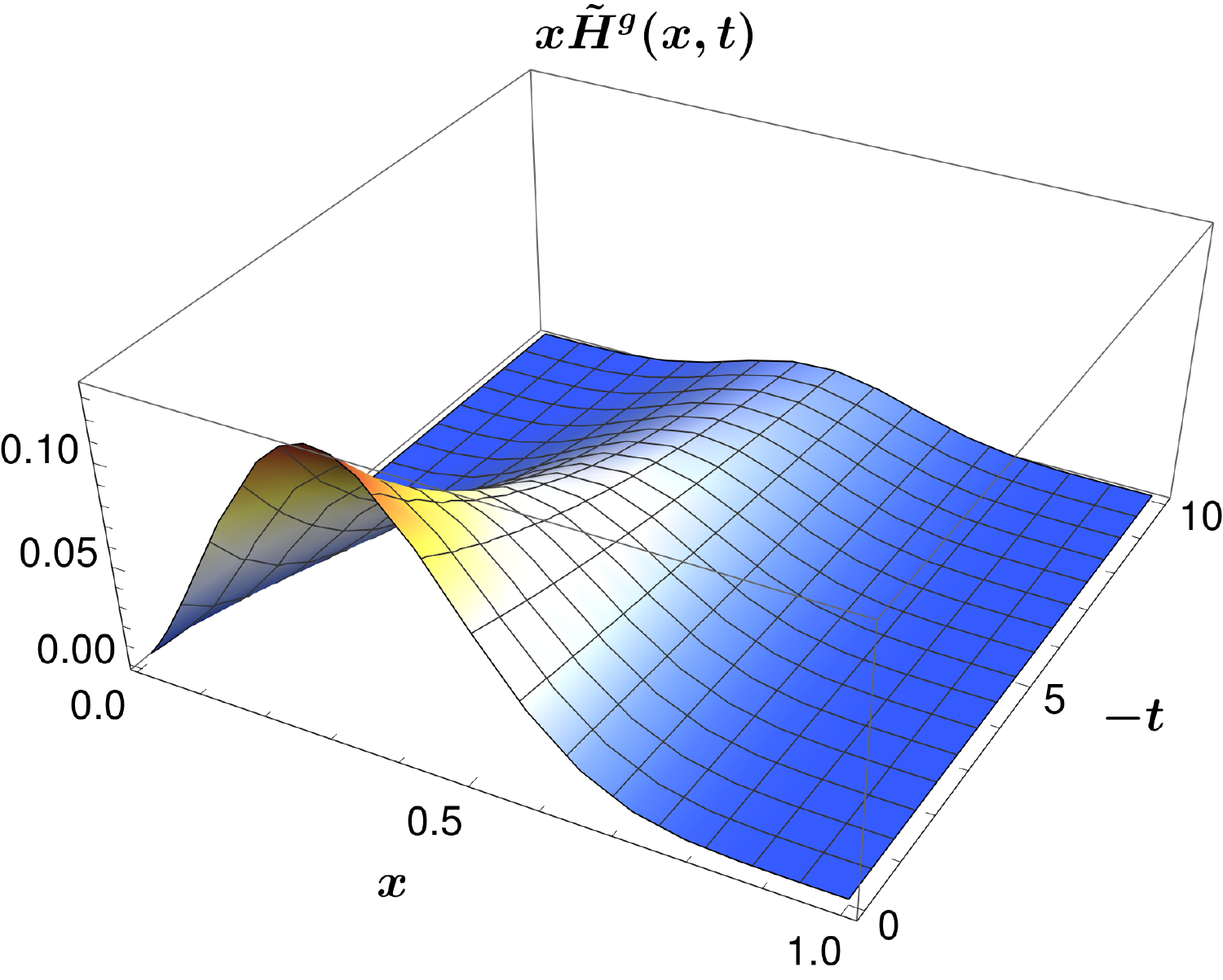}
\end{center}
\caption{Three-dimensional structure of the unpolarized (upper panel) and polarized (lower panel) gluon GPDs in the proton as function of $x$ and $-t$ (in units of GeV$^2$). These results are generated using the Pomeron exchange with scale parameter $\lambda_g = 1/4  \alpha_P^\prime \simeq 1 \, {\rm GeV}^2$. The intercepts of the Pomeron trajectories associated with the unpolarized and the polarized GPDs are $\alpha_P(0)=1.08$ and $\widetilde{\alpha}_P(0)=0.08$, respectively.}
\label{gluonGPDs}
\end{figure}
\subsection{Gluon GPDs}\label{gluonprotonGPDs}
Using the expressions of the gluon GFF in Eqs.~\eqref{GFF1} and \eqref{GPDH}, we write the upolarized gluon GPDs at skewness zero in the proton, choosing specific $x$ and $t$ dependences of the GPDs~\cite{deTeramond:2018ecg} as:
\be  \label{unpGPD}
H_{\tau}^{g}(x,t)=g_{\tau}(x)\,e^{tf(x)}\,,
\ee
where the unpolarized gluon PDF of twist-$\tau$, $g_{\tau}(x)$, and the universal profile function $f(x)$ are given in Eqs.~\eqref{gtaux} and \eqref{profilefun}, respectively. In a similar fashion, we express the polarized gluon GPD 
\be  \label{helicityGPD}
\tilde{H}_{\tau}^{g}(x,t)=\Delta g_{\tau}(x)\,e^{tf(x)},
\ee
with the polarized gluon PDF defined in Eq.~\eqref{helicitypdf}. Note that we consider the same $t$-dependence factor in both the GPDs. This emerges from the
linear Regge trajectories associated with the unpolarized and polarized distribution having
equal slope. Meanwhile, different intercepts of the trajectories generate different $x$-dependence
structure of those GPDs.

The three-dimensional structures of  the gluon GPDs as function of $x$ and $-t=Q^2=\vec{q}_\perp^2$ are illustrated in Fig.~\ref{gluonGPDs}. In the forward limit, $-t=0$, the GPDs reduce to their corresponding collinear PDFs. The unpolarized gluon distribution peaks at small-$x$, while the polarized GPD has its peak located slightly higher value of $x$ than the unpolarized GPD. The magnitude of $xH^{g}(x,t)$ is much higher than that of $x\tilde{H}^g(x,t)$. The peaks of these GPDs move toward higher values of $x$ and simultaneously reduce the magnitudes with increasing the value of the momentum transfer $-t$. This seems to be a model
independent behavior of the GPDs, which has also been observed in quark GPDs evaluated within this HLFQCD framework~\cite{deTeramond:2018ecg} as well as in various phenomenological models for the nucleon~\cite{Ji:1997gm,Scopetta:2002xq,Boffi:2002yy,Boffi:2003yj,Vega:2010ns,Chakrabarti:2013gra,Mondal:2015uha,Chakrabarti:2015ama,Mondal:2017wbf,Xu:2021wwj,Kriesten:2021sqc}. 
As gluon GPDs are not yet experimentally determined, it is not possible to compare  our predictions with any data. Nonetheless,  
the gluon GPDs can be investigated experimentally from Deeply Virtual Compton Scattering (DVCS) and other exclusive processes. The upcoming EICs in the USA~\cite{AbdulKhalek:2021gbh} and in China~\cite{Anderle:2021wcy} can significantly improve our current knowledge of the gluon GPDs.
Simulation studies in Ref.~\cite{Aschenauer:2013hhw} showed that the proposed high-luminosity EICs can perform accurate measurements of DVCS cross sections and asymmetries in a very fine binning and with a very low statistical uncertainty.

The GPDs in transverse impact parameter space are obtained via the Fourier transform of the GPDs with respect to the momentum transfer along the transverse direction $\vec{q}_\perp$~\cite{Burkardt:2002hr}:
\begin{align}
 \mathcal{F}(x, {b}_\perp)& =
\int \frac{{\rm d}^2{\vec q}_\perp}{(2\pi)^2}
e^{-i {\vec q}_\perp \cdot {\vec b}_\perp }
F(x,0,t)\,,\label{eq:Hb}
\end{align}
with $F$ being the GPDs in momentum space and $\vec{b}_{\perp}$ defines the transverse impact parameter conjugate to the transverse momentum transfer $\vec{q}_{\perp}$. The function $\mathcal{H}^g(x, b_{\perp})$ can be interpreted as the number density of gluon with longitudinal momentum fraction $x$ at a given transverse distance $b_{\perp}$ in the proton~\cite{Burkardt:2000za}. We can then define the $x$-dependent squared radius of the gluon density in the transverse plane as~\cite{Dupre:2016mai}:
\begin{align}
\langle b^2_\perp \rangle^g (x) = \frac{ \int {\rm d}^2 {\vec{ b}_\perp} { b}^2_\perp \mathcal{H}^g(x, {b_{\perp}})}{\int {\rm d}^2 {\vec{ b}_\perp}  \mathcal{H}^g(x, b_{\perp})},
\label{eq:cr5}
\end{align}
which is uniquely determined by the profile function $f(x)$, Eq.~\eqref{profilefun}. 
\begin{figure}
\begin{center}
	\includegraphics[scale=0.5]{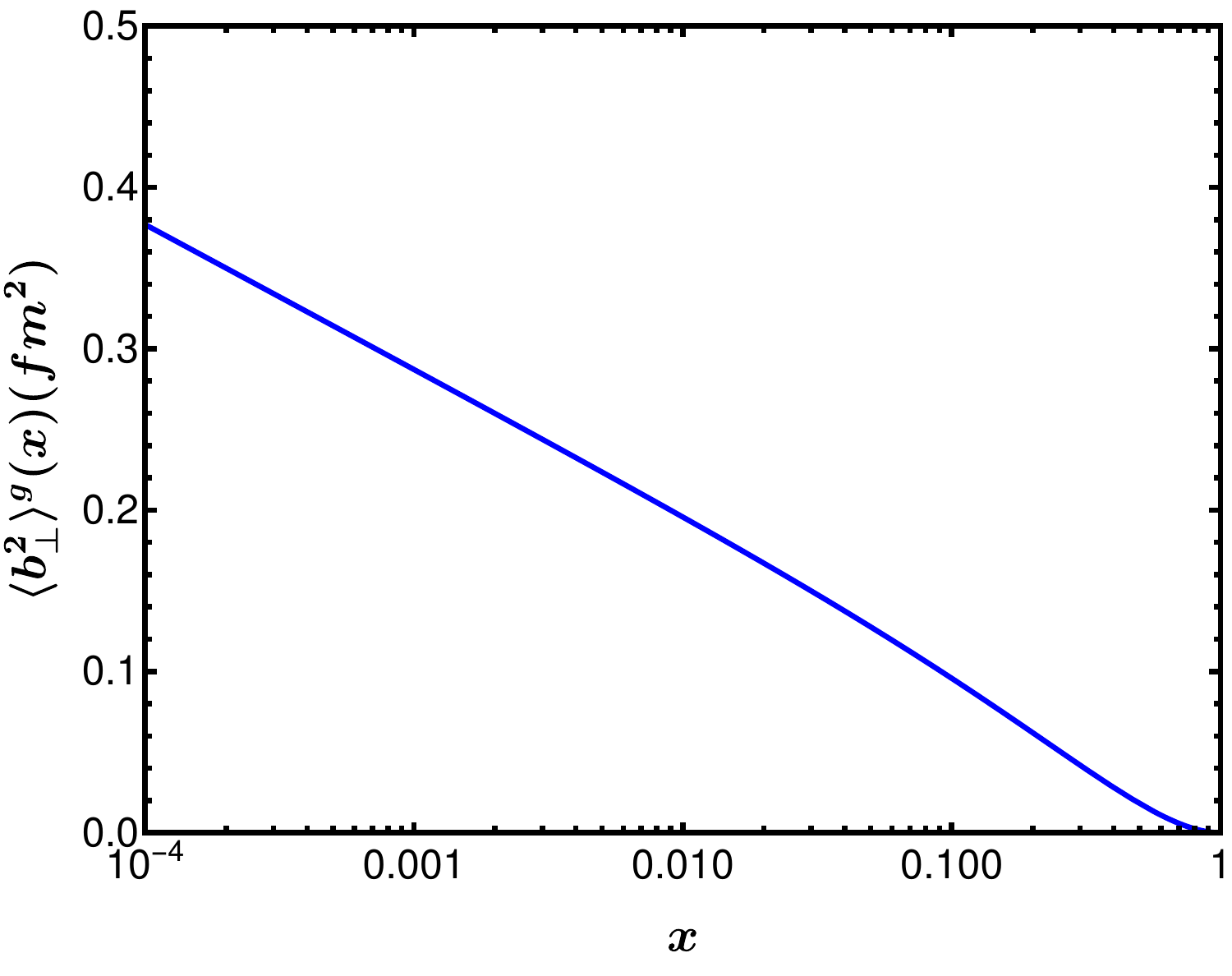}
\end{center}
	\caption{$x$-dependence of $\langle b_{\perp}^{2}\rangle $ for gluon in the proton. This result is obtained using the Pomeron exchange with scale parameter $\lambda_g = 1/4  \alpha_P^\prime \simeq 1 \, {\rm GeV}^2$.}
	\label{b2x}
\end{figure}
We present the $x$-dependent squared radius of the proton's gluon distribution in Fig.~\ref{b2x}. It shows an increase of transverse radius with decreasing value of the gluon momentum fraction $x$. At large-$x$, the
transverse size of the distribution behaves as a point-like color-singlet object. This nature is the
origin of color transparency in nuclei~\cite{Brodsky:2022bum}. Note that this behavior is universal and depends only on the profile function $f(x)$, which, in LFHQCD,
is determined by the hadron mass scale $\lambda_g$ and the universal reparametrization function $w(x)$. The general features of $\langle b^2_\perp \rangle^g (x)$ as reported here have also been observed in the dependence of the transverse size of the proton on the quark's longitudinal momentum, which has been determined from DVCS experimental data~\cite{Dupre:2016mai} and investigated in other theoretical studies~\cite{Xu:2021wwj, Brodsky:2022bum}.

\section{conclusion}\label{discussion_conclusion}
 We have evaluated the polarized gluon distribution using a unified nonperturbative approach based on the gauge-gravity correspondence, light-front holography and the generalized Veneziano model. The gluon PDFs can be expressed in terms of a universal reparametrization function $w(x)$~\cite{deTeramond:2018ecg,Liu:2019vsn,deTeramond:2021lxc}. A simple ansatz for $w(x)$, which satisfies the pQCD constraints for the gluon helicity asymmetry ratio at the end points, $x\to \{0,\,1\}$, leads to a precise description of gluon helicity distribution. We have observed a good consistency between
our prediction for the gluon helicity PDF and the global fits as well as with the results obtained from various theoretical approaches. The gluon helicity asymmetry ratio is found to be in good agreement with the COMPASS data. Within the HLFQCD framework, we have predicted that at the scale $\mu^2\sim 1$ GeV$^2$, the gluon helicity  contributes
\be
\Delta G=0.221^{+0.056}_{-0.044}\,,
\ee
which is almost $44\%$ of the proton spin.  Experimentally, there still remain large uncertainties about the small-$x$
contribution to $\Delta G$. Precise determinations of the gluon helicity distribution in the $x<0.02$ region are required to constrain $\Delta G$.

We have subsequently presented the unpolarized and polarized gluon GPDs in the proton, choosing specific $x$ and $t$ dependences of the GPDs, using the gluon PDFs and the universal profile function $f(x)$, which can also be expressed in terms of the universal reparametrization function $w(x)$. We have
observed that the unpolarized gluon GPD is distinctly different from the gluon helicity dependent GPD,
whereas the difference between them is given by their corresponding collinear PDFs. We have found that the qualitative behavior of the GPDs in HLFQCD  approach bears similarities to other phenomenological models.

\begin{acknowledgments}
{\it acknowledgments}.---CM is supported by new faculty start up funding by the Institute of Modern Physics, Chinese Academy of Sciences, Grant No. E129952YR0.  CM also thanks the Chinese Academy of Sciences Presidents International Fellowship Initiative for the support via Grants No. 2021PM0023. The work of DC is supported by Science and Engineering Research Board under the Grant No. CRG/2019/000895.
\end{acknowledgments}

\bibliography{references.bib}
\end{document}